# THE CONTROL SYSTEM OF THE ELLIPTICAL CAVITY AND CRYOMODULE TEST STAND DEMONSTRATOR FOR ESS

A. Gaget†, T. Joannem†,Q. Bertrand, A. Gomes, J.F Lecointe, D. Loiseau, IRFU, CEA, Université Paris-Saclay, F-91191, Gif-sur-Yvette, France

## Abstract

CEA IRFU Saclay is taking part of ESS (European Spallation Source) construction through several packages and, especially in the last three years on the Elliptical Cavity and Cryomodule Test Stand Demonstrator (ECCTD). The project consists of RF test, conditioning, cryogenic cooldown and regulations of eight cryomodules with theirs four cavities each. For now, two medium beta cavities cryomodules have been successfully tested. This paper describes the context and the realization of the control system for cryogenic and RF processes, added to cavities tuning motorization relying on COTS solutions: Siemens PLC, EtherCAT Beckhoff modules, IOxOS fast acquisition cards and MRF timing cards.

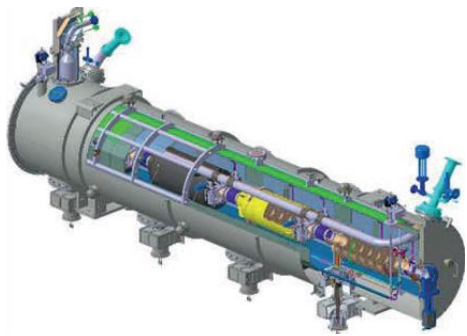

Figure 1: Medium-beta cryomodule general layout.

## INTRODUCTION

The European Spallation Source ESS is a large European research infrastructure under construction in Lund, Sweden. It will be composed among others of 30 elliptical cryomodules. These will be integrated in the next few years with a delivery rate of one cryomodule per month. An international collaboration has been established to develop and construct the 30 elliptical cryomodules. CEA Saclay and IPN Orsay are collaborating to design, build and test a first Medium beta Elliptical Cavities Cryomodule Demonstrator (M-ECCTD) [1] (see Figure 1). A second demonstrator with high beta cavities (H-ECCTD) is being developed by CEA before starting the production of the cryomodules. CEA will provide many other components, such as the power couplers and their RF processing.

_______________________
† alexis.gaget@cea.fr
† tom.joannem@cea.fr

The control for the high level control activities has been developed in EPICS, based on the ESS EPICS Environment (EEE) [2] developed by the ESS control team at Lund (ICS). For the critical needs (cryogenic, vacuum, interlock), we have used Siemens PLC with an homemade SCADA dedicated to those devices and communicating via TCP/IP: Muscade®.

## CONTEXT

Before the cryomodule package the CEA has developed a platform for our previous contribution, the conditioning of the power couplers for the cavities. To fulfill this need, it has been decided to upgrade an existing 704MHz RF platform of our own. EPICS has replaced all the LabVIEW software that was controlling the RF source. This RF Source is now used intermittently to condition power couplers that will be used in the cryomodule, and to condition cavities in the cryomodule demonstrator as shown in Figure 2.

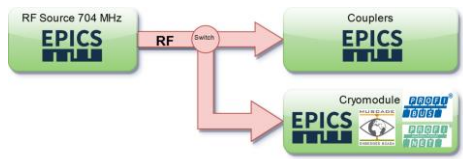

Figure 2: RF platform at CEA.

Likewise an existing cryogenic platform called Supratech is used to provide cryogenic fluids like Helium and Nitrogen to local experiments. In order to simulate the valve box for cryomodules cooling down an intermediate gateway control system was created.

Supratech control system is based on Siemens PLC and Profinet fieldbus to communicate. It was decided to follow the same logic for the cryomodule and the cryogenic gateway control systems.

Thanks to this homogeneity and more precisely Profinet, cryomodule control system can easily share data with other systems and send commands directly to the local cryoplant to refill it with cryogenic fluids.

## HARDWARE SOLUTION

### Fast Acquisition

For RF signal, photomultiplier and pickup electron measurement, we needed acquisition cards able to acquire data with a sampling rate up to 1Ms/s and with a frequency period up to 14Hz. Today ESS and CEA have now both evolved to MTCA standard but at the project beginning we chose the proven VME solution. We are using IOxOS [3] hardware solution with the VME64X CPU card IFC1210 coupled with the ADC3111 FMC (FPGA Mezzanine Card) (see Figure 3). ESS ICS have developed the drivers.

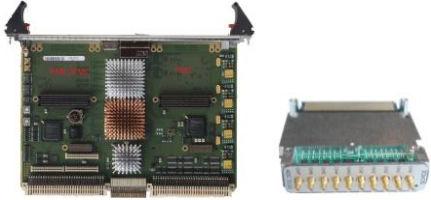

Figure 3: IOxOS cards (IFC1210 and ADC3111).

The ADC3111 is an FMC mezzanine acquisition board, which allows acquiring up to 8 channels at a sampling rate of up to 250MHz.

The VME64X CPU IFC1210 is a single board computer on which a real-time Linux kernel runs and allows plugging two FMC cards or one FMC and one PMC (PCI Mezzanine Card) as shown in Figure 4.

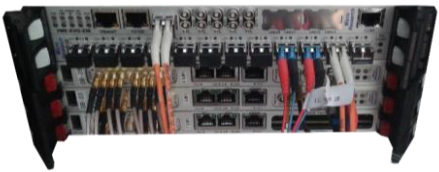

Figure 4: Example of VME rack used.

## Timing System

The timing system based on MRF Timing system [4] here is quite simple as it is only use to produce synchronized RF pulse, High voltage, piezoelectric motor and acquisition trigger. The MRF Timing system consists of an Event Generator (EVG) that converts timing events and signals to an optical signal distributed through Fan-Out Units to an array of Event Receivers (EVRs) (see Figure 5). The EVRs decode the optical signal and produce hardware and software output signals based on the timing events received.

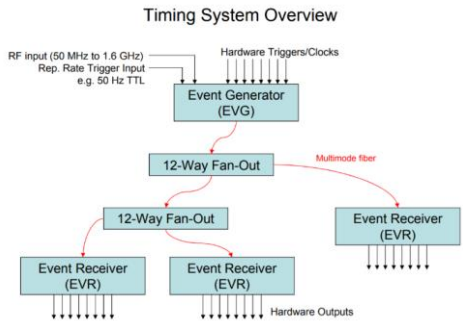

Figure 5: Timing System Overview [5].

The MRF timing system, that we have chosen, can appear oversized for a test stand but the possibility to diffuse simply the signal through optical fiber have quickly shown its importance and the flexibility to link future other test stands should suit our needs.

## PLC Hardware

PLC technology is based on Siemens PLC 1500 generation [6]. CPU cards 1516-3 PN/DP (see Figure 6) were chosen because of their network flexibility thanks to two Profinet/TCP-IP and one Profibus DP embedded cards. Input and output cards of the same generation are used for global homogeneity and for their flexibility. For example: analog input cards of eight channels allow versatile acquisition range: from +-50mV to +-10V, 4-20mA and PT100 measurement. Extension racks are communicating with CPU using Siemens native fieldbus Profinet. An exception is made for Phytron motorization only supported by ET200S from Siemens PLC 300 generation.

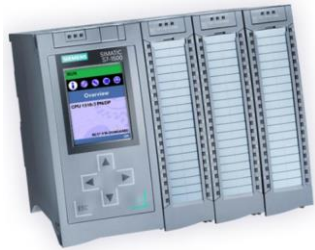

Figure 6: Siemens 1516-3 PN/DP CPU and cards.

## Slow Acquisition

In addition, when the PLC is not required we have used the Beckhoff EtherCAT [7] modules using EtherLab [8] driver from IgH to be able to use these modules through EPICS.

# SOFTWARE SOLUTION

## EPICS

The distributed control system is based on EPICS V3.15.4. For the environment of development, we have used the ESS EPICS Environment (EEE) V2.1.1, ESS ICS has now a new environment (E3) and although EPICS modules developed are compatible with it, the first test stand has been in production for 3 years so for homogeneity it has been decided to keep using this environment.

For GUI we have used the ESS version of CS-Studio. You can see an example of a view for a cavity in the Fig 7.

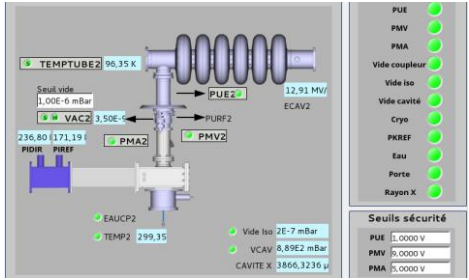

Figure 7: Cavities GUI.

## TIA PORTAL

On many experiments, PLC developments give a huge library of standard functions to the laboratory. Adding to these functions, PLC code is following quality requirements and a similar architecture. This architecture was designed to make the maintenance as easy as possible for every PLC expert using TIA Portal V15.1.

The languages used are PLC worldwide standards: LADDER, SFC and SCL.

The homemade CEA SCADA called Muscade® is used on the cryomodule test bench.

The Figure 8 shows the cryogenic view of the cryomodule test bench. This view is dedicated to PLC sensors, actuators and automatic procedures. Temperature, pressure and flow sensors are available from this view. Automatic procedure for cooling down, motorization and regulations are also available from the right of the view.

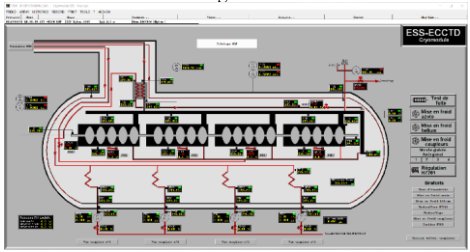

Figure 8: Cryomodule cryogenic PLC.

## Gateway EPICS / PLC

To communicate between EPICS and PLCs the use of the S7plc driver [9] has been decided.

## COUPLER CONDITIONING

A conditioning of the couplers consists in cleaning it by rising RF power in the couplers with different parameters like duty factors, intensity, frequency. You can see an example shown in the Figure 9. Couplers have already been conditioned on their own test bench [10] once. During this test, they will be warm and cold conditioned, these two operations are made automatically by the control system.

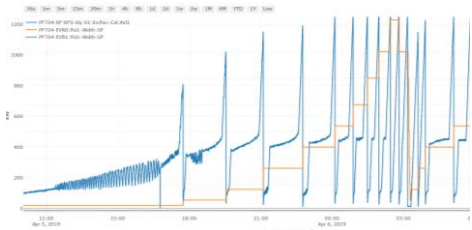

Figure 9: Complete conditioning sequence, plotting RF power vs. time.

Considering an important amount of similar development, we have developed an EPICS modules toolkit for RF test stands [11, 12]. This toolkit allows executing a long succession of conditioning sequences predefined and configured by the user.

The operator configures several conditioning sequences and thanks to another mechanism, can configure a "master sequence" to chain RF ramp sequences as shown in Figure 10. This method allows the operator to launch automatically many sequences in one shot. In case of problems, he will be warned by phone through the BEAST Alarm.

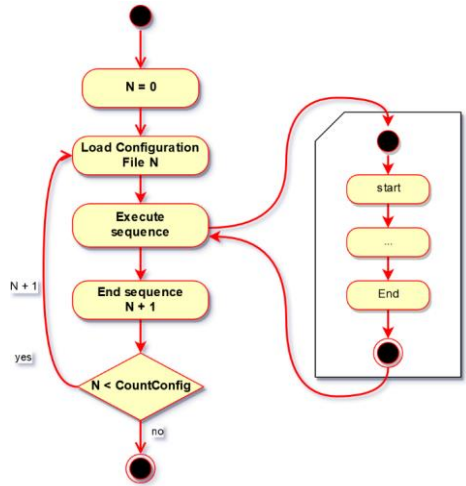

Figure 10: Load restore sequence.

This toolkit has already been tested and used for the conditioning of the power couplers for the cavities.

## CAVITY TUNING SYSTEM

Depending on their temperature, cavities due to their composition can be deformed. To overcome the problem we use motor to adjust the cavity shape, and so reach the target frequency. It is called the cavity tuning. It uses two systems: Phytron and piezoelectric motorizations. These

Commenté [GF1]: ???

two systems take in charge different tasks. First, the Phytron motor has to do a definite movement and then stays static. Secondly, when the Phytron motor reaches its position, the piezoelectric motorization control has to do a dynamic regulation at a high frequency up to 14Hz in the case of ESS cavities. These two systems use different technologies but work together.

*Phytron*

To be controlled the Phytron motorization has a dedicated Siemens controller. This controller is hosted on a Siemens ET200S of the 300 generation rack. The control uses an open loop. A step-by-step control is done to obtain a good position-frequency correlation. The control loop is open because of technologic restrictions at the time of the design: no position sensor was able to work in cryogenic and vacuum areas. For the last control prototypes, a new generation position sensor has been added in order to have a closed control loop and to increase precision and security. The Figure 11 represents the GUI used by an operator to control the Phytron motor.

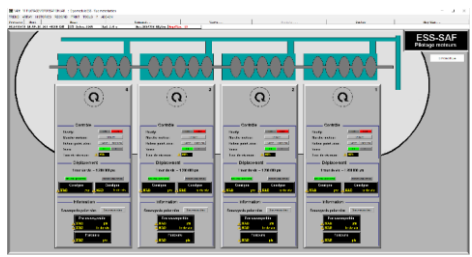

Figure 11: PLC Phytron control view.

*Piezoelectric Motor*

As shown in Figure 12 the piezoelectric motor is controlled through a function generator AFG3252C. Synchronized to the RF pulse, the generator sends a signal shape configured by the operator. Then this driver applies voltage to the Motor.

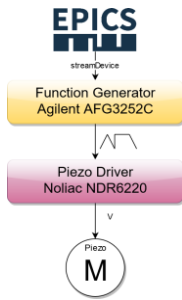

Figure 12: Piezoelectric operation architecture.

This generator is controlled with EPICS and Streamdevice driver that allows using SCPI commands. The Figure 13 is showing the graphical interface used by the operator to configure and compute the function for each piezoelectric motor.

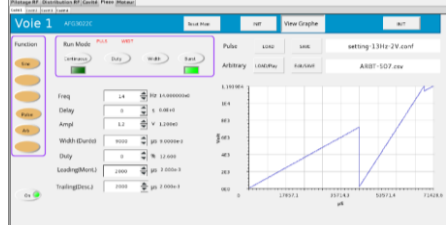

Figure 13: Generator function AFG3000 GUI.

*Automatic Cavity Tuning Characterization*

For the first cryomodule tested, operators made the characterization of the cavity tuning manually. As they were using the network analyzer Agilent E5060A that communicates through TCP/IP, it has been decided to use EPICS to control this equipment. EPICS is also used to unify information coming from the analyzer and PLC, and so makes easier to develop an automatic sequence for the measurement.

The sequence has been implemented with the SNL sequencer. It consists, as shown in Fig. 14, in moving a Phytron motor with a defined movement in one direction and applying different SPAN on the measure. During all the sequence, the centroid is archived in the EPICS archive appliance and with this value, we can characterize the cavity tuning and check that the hysteresis is correct.

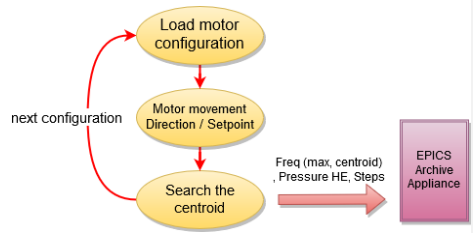

Figure 14: Sequence for the automatic cavity tuning measurement.

# CRYOGENIC CONTROL

Cryomodules have to be cool down with Nitrogen and Helium provided by a local cryoplant. This cryoplant and the cryomodule test bench use Siemens PLC from different generations. A Profinet communication was established in order to communicate between those two systems (see Figure 15).

Cryofluids are monitored via temperature, pressure and flow sensors. Cryogenic temperature sensors are connected to a CEA homemade conditioning device called Boranet communicating through Profinet.

Analog cryogenic valves for cryofluids have remote actuators to be as similar as possible to the ESS Lund cry-omodules configuration. Those remote actuators are Siemens remote Sipart PS2 and can control remote actuators through 4-20mA signals. In fact, electronic elements are located into a cabinet outside of the cryomodule bunker and inside the bunker actuators are without electronic. With this technology, we respect the ESS non-electronic device requirement inside the tunnel.

With data received by cryogenic sensors and commands sent to cryogenic actuators, automatic cooling procedures are reliable and work during months without any interruption.

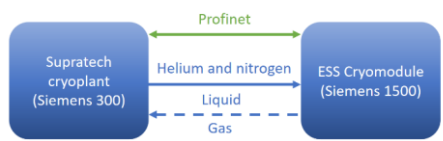

Figure 15: Control and cryogenic exchanges between Supratech cryoplant and ESS cryomodule.

## VACUUM CONTROL

In order to obtain a high quality vacuum treatment for cavities in the clean room, dedicated pumping groups were developed. These groups can be moved to different clean room slots, make slow pumping in order to avoid generation and movement of particles, and run residual gas analysis.

A Siemens PLC is embedded in each group and controls every device, including residual gas analysis. This communication through TCP-IP protocol gives the possibility having full automatic procedures to obtain automatic pumping procedures, diagnostics, leak check and residual gas analysis on a single interface.

This innovative architecture, by using a mobile and plug and play graphic interface gives the possibility, to be autonomous and manage up to eight groups in parallel lines with only two interfaces.

## FUTURE APPLICATIONS

As mentioned before, the knowledge obtained by creating the ESS cryomodule test bench control system will help us a lot for following projects, like SARAF cryomodule test bench. In fact, these two test benches will be quite similar. This experience feedback will be integrated into the SARAF test stand.

Some significant evolutions will be integrated:

- Cryogenic and vacuum compliant position resolver for a close loop regulation of Phytron motors for cavity tuning will be added for a better security
- VME boards will be replaced by MTCA boards but as we keep the same mezzanine cards, standard EPICS development still could be reused.
- Boranet is no longer available and new CEA development named CABTR [13], used on ITER, will be used on the next test bench.

- The CEA homemade SCADA Muscade® will no longer be used, it has been decided to migrate to EPICS and we have developed some tools to have functionalities that does not exist in EPICS [14].
- Learning from ESS EEE, we have created our own environment of development called IEE (Irfu EPICS Environment) [15] that corresponds better to our standards..

## CONCLUSION

The control system of the ESS cryomodule test bench is now fully functional. It has been a rewarding challenge as it brings us a lot of experience and allows us to develop some real interesting tools (pumping groups, EPICS toolkit …). Furthermore, it will be very useful for other projects (SARAF, IFMIF).

## ACKNOWLEDGEMENT

The authors would like to thank O. Piquet, T. Hamelin, G. Devanz, P. Sahuquet and F. Gouit that have been patient users of our system and all the people that contributed to the installation and conception of the cryomodule test stand.